\documentclass[prd,aps,nofootinbib]{revtex4}
\usepackage{graphicx}
\usepackage{bbold}
\usepackage{slashed}
\usepackage{amsmath}
\usepackage{amssymb}
\usepackage{moresize}
\usepackage{float}
\restylefloat{table}
\usepackage{tikz} 
\usepackage[colorinlistoftodos]{todonotes}
\usepackage[toc]{appendix}
\usepackage{extpfeil}

\newcommand{\bea}{\begin{eqnarray}}
\newcommand{\eea}{\end{eqnarray}}
\newcommand{\be}{\begin{equation}}
\newcommand{\ee}{\end{equation}}

\newcommand{\bal}{\begin{align}}
\newcommand{\eal}{\end{align}}
\newcommand{\cm}{\color{black}}

\newcommand{\cb}{\color{black}}

\newcommand{\cbl}{\color{black}}
\newcommand{\bL}{\begin{Large}}
\newcommand{\eL}{\end{Large}}

\DeclareMathOperator{\sech}{sech}
\DeclareMathOperator{\cosech}{cosech}

\newcommand{\cP}{\ensuremath{\mathcal{P}}}
\newcommand{\cT}{\ensuremath{\mathcal{T}}}
\newcommand{\cC}{\ensuremath{\mathcal{C}}}
\newcommand{\cPT}{\ensuremath{\mathcal{PT}}}
\newcommand{\cCPT}{\ensuremath{\mathcal{CPT}}}

\begin{document}

\preprint[{\leftline{KCL-PH-TH/2024-{\bf 36}}

\noindent
\title{$\cPT$ symmetric fermionic particle oscillations in even dimensional representations }
\author{Leqian Chen
}
\author{Sarben Sarkar
}
\affiliation{
Theoretical Particle Physics and Cosmology, King's College London, Strand, London, WC2R 2LS, UK}
\begin{abstract}
We describe a novel class of quantum mechanical particle oscillations in both relativistic and nonrelativistic  systems  based on $\cPT$  symmetry and $\cT^2=-1$ (\cb relevant for fermions\cbl), where $\cP$ is parity and $\cT$ is time reversal. The Hamiltonians are chosen at the outset to be self-adjoint with respect to a \cPT~inner product. The quantum mechanical time evolution is based on a modified $\cCPT$  inner product  constructed in terms of a suitable $\cC$ operator. The resulting quantum mechanical evolution is shown to be unitary and probability is conserved  by the oscillations.
\end{abstract}

\keywords{$\mathcal{PT}$ symmetry \sep quantum mechanics \sep path integral \sep epsilon expansion \sep renormalisation group \sep non-Hermiticity \sep axion}

\maketitle

\section{Introduction} 
$\cPT$-symmetric quantum systems \cite{Bender1999-bh,Bender:2007nj,R2,Christodoulides:2018arz,Moiseyev2011-cr} offer a new paradigm for constructing non-Hermitian physical theories.  They have a discrete $ \cPT$ symmetry \cite{R1,R2},  where  $\cP$ is a linear operator such as parity and $\cT$ is an  antilinear operator such as time reversal \cite{Wigner2012-lr,Messiah2014-ap,Domingos1979-mw}. Such quantum systems \cb are candidates for theories of fundamental physics,  when \cPT ~symmetry is unbroken i.e.  \cbl when  observables  have real eigenvalues \cite{Bender1999-bh,R16} and time evolution  is unitary \cite{Weigert:2003py,Das:2009it,Bender:2003gu,Bender:2003fi}\footnote{\cb  There are important applications of broken \cPT ~symmetry in condensed matter and photonics \cite{Christodoulides:2018arz,Feng2017-ph,Ozdemir2019-jf,Miri2019-bl,Bergholtz2021-lx,krasnok2021paritytimesymmetryexceptionalpoints} as well as in quantum chromodynamics \cite{Schindler:2021otf,Schindler:2019ugo}. \cbl}. This last requirement is essential \cb for unitarity \cbl and involves the construction of an inner product on the Hilbert space, which differs from the Dirac inner product used in  Hermitian theories. The explicit construction of inner products is a non-trivial issue for the study of particle oscillation phenomena \cite{Alexandre:2023afi,Ohlsson:2019noy,Ohlsson:2015xsa}. \cb Neutrino oscillations (discussed in Section \ref{compar} and Appendix \ref{NOscill}  )  is one of the most significant discoveries \cite{Athar2022-ud,Huber:2022lpm} in particle physics, providing strong evidence for physics beyond the Standard Model (BSM). It is interesting to consider the implication of $\cPT$ symmetry in providing a new paradigm for fermionic flavour oscillations, where the dispersion relations do not depend on individual masses of fermions, required in conventional descriptions.\cbl
\vskip .1cm
 In the search for theories for BSM   \cite{Ramond2003-jv,Nagashima:2014tva,Burgess2011-vc}, it is suggested that non-Hermitian, but $\cPT$-symmetric, theories \cite{R3.1, R3.2, R3.3, R3.4, R3.5, R3.6, R3.7, R3.8, R3.9, R3.10, R3.11, R3.12, R3.13, R3.14, R3.15} can play a role. This is an important motivation since \cPT~symmetry enhances the set of theories, which are physically interesting\footnote{Moreover, starting from Hermitian theories, there are instances where the effects of renormalisation lead to non-Hermitian features and a \cPT-symmetric interpretation allows the avoidance of ghosts \cite{R5,Croney:2023gwy,Lencses:2022ira}}  . \cPT-symmetric theories split into the two classes $ \cT^{2}=\pm 1 $ \cite{PhysRevA.82.042101}. The case of  bosonic systems, where $ \cT^{2}=1$ \cite{Deng:2012yw},  has been discussed widely (see for example \cite{R2,Christodoulides:2018arz,Moiseyev2011-cr}).  Fermionic systems belong to the less widely studied class  $ \cT^{2}=-1$ \cite{PhysRevA.82.042101,Jones-Smith:2009qeu,Beygi_2019,Cherbal:2012pm,Croney:2023gwy,R3a}. We only consider the quantum mechanical Hilbert space of the $\cT^{2}=-1$ case for \emph{both} a non-relativistic and a relativistic fermion, where the Hamiltonian has a  matrix structure.~\cb Since we are interested in fermions, which could be  neutrinos, it is the $\cT^{2}=-1$ case that is appropriate. Our study  is based on a detailed construction of a positive definite inner product in the case of unbroken $\cPT$ symmetry, and this gives a description of oscillations which conserves probability. We compare the expression of our oscillation formula with that from the  conventional oscillation formula.  \cbl
\vskip .2cm
For our models we define suitable flavour states and study flavour oscillation phenomena that arise, which differ  from those in the $\cT^{2}=1$ case \cite{Ohlsson:2019noy,Ohlsson:2015xsa}. The models are based at the \emph{outset} on an intrinsically  $\cPT$-symmetric (possibly relativistic) theory requiring \emph{self-adjointness} of operators with respect to an \emph{inner product} distinct from the conventional Dirac inner product \cite{Sablevice:2023odu}. For the class $\cT^{2}=-1$, the study of theories based on the principles of \cPT~ self-adjointness  was  initiated  in \cite{PhysRevA.82.042101,Jones-Smith:2009qeu,JonesSmith2010a} for necessarily even dimensional representations of the wave function. This investigation represents a first step in  constructing a new class of field theories for the $\cT^{2}=-1$ case (which could be relevant for BSM). \cb Hence, the framework of the studies \cite{PhysRevA.82.042101,Jones-Smith:2009qeu,JonesSmith2010a} moves away from perturbing a Hamiltonian with a special type of mass matrix, and herein lies their significance. The Hamiltonian itself is determined by the self-consistency requirements of the formalism.~\cbl Even a path integral approach to the construction of \cPT-symmetric quantum field theories \cite{Croney:2023gwy,R3a} relies implcitly  on the existence of an underlying Hilbert space and inner product.\footnote{We do not discuss relativistic field theory, which has been considered recently by us in a path integral formulation. Although the Hilbert space inner product plays a role in the path integral measure, it is still possible to derive Schwinger-Dyson equations without a detailed construction of the inner product \cite{R19,Jones:2009br}.}. 

\vskip .2cm
The first discussion  of $\cPT$-symmetric quantum mechanics for $ \cT^{2}=-1$ (which is denoted $\cT$ odd) is given in \cite{PhysRevA.82.042101,Jones-Smith:2009qeu}. In these works a new inner product (based on  earlier work \cite{R16} for the case $ \cT^{2}=1$ ), is proposed. This inner product incorporates  a different representation for the  $\cT$ operator, which is antilinear \cite{Wigner2012-lr}. The $\cT$ operator includes, apart from complex conjugation,  a linear operator $Z$, i.e. $\cT \psi =Z\hat{K}\psi $ where $\hat{K}$ is the complex conjugation operator and $\psi$ is any state in the Hilbert space. An example of such a $\cT$ operator is $\cT=i\gamma^{1} \gamma^{3}\hat{K} $ for the Dirac equation \cite{R20} (in terms of  gamma matrices in the Dirac representation).
It is necessary to investigate that the proposed inner product leads to a unitary theory, i.e.    probability is conserved for non-Hermitian but \cPT-symmetric Hamiltonians. We precisely investigate this for the case of flavour oscillations in  models, which involve  matrix representations of the Hamiltonian. These models have intrinsic interest, but also allow an explicit construction of an inner product\footnote{Generally a closed form solution for the inner product is \emph{not} possible.There is also no proof that there is a unique inner product.} and, hence, an investigation of \emph{unitarity} in the context of a new type of flavour oscillation. 
\vskip .3cm
  A common approach to the construction of $\cPT$-symmetric Hamiltonians is to start from a Hermitian  Hamiltonian to which a \cPT -symmetric (but non-Hermitian) piece is added. We follow the more radical approach \cite{Jones-Smith:2009qeu,Sablevice:2023odu}  of constructing, at the outset, a $\cPT$-symmetric (possibly relativistic) theory requiring \emph{self-adjointness} of operators with respect to an \emph{inner product} different from the conventional Dirac inner product \cite{Sablevice:2023odu}. This program,  \cite{PhysRevA.82.042101,Jones-Smith:2009qeu,JonesSmith2010a} led to a study of even dimensional representations of the wave function. In the relativistic case  too, the use of a non-Hermitan representation of the Clifford algebra  leads to even dimensional representations of the wavefunction \cite{Jones-Smith:2009qeu}. The use of higher dimensional wave functions allows  unconventional flavour-like oscillations also in the relativistic case. Our key focus is establishing conditions for unitary evolution  in intrinsically $\cPT$-symmetric $\cT$-odd Hamiltonians.  The construction of inner products and the related notion of pseudo-Hermiticity \cite{R16,Mostafazadeh_2005,Mostafazadeh:2005wm,Schulze-Halberg2017-lm,Bender_2006,Das_2011,Zhang:2019gyc} is extended to our \cT-odd case. 
  \vskip .3cm
 The plan of this paper is as follows:
\begin{enumerate}
\item We  discuss  inner products for even-dimensional wave functions arising in  $\cT^2=-1$,  $\cPT$-symmetric quantum mechanical theories and relate them to pseudo-Hermiticity.
\item The formalism is extended to the relativistic wave equation of Dirac;  a new modified inner product is introduced, which differs from that in \cite{PhysRevA.82.042101,Jones-Smith:2009qeu}. However, the consistency conditions required for a non-Hermitian representation of the Clifford algebra, found in \cite{PhysRevA.82.042101,Jones-Smith:2009qeu}, remain unchanged.
\item A consistent $\cPT$-invariant four dimensional matrix model is constructed with an explicit $\cCPT$ inner product in the Hilbert space, which leads to identical consistency  conditions found in \cite{Jones-Smith:2009qeu}.~This formalism is applied to a type of  flavour oscillations.
\item A relativistic eight-dimensional representation of the Dirac equation is constructed, which is non-Hermitian but \cPT-symmetric. The use of the new $\cPT$ inner product, referred to above, allows the construction of a  $\cCPT$ inner product, where $\cC$ is a hidden symmetry \cite{Bender:2004guu} of the Hamiltonian \footnote{We should stress that $\cC$ is not the charge conjugation operator of the Dirac equation.}. A new form of probability-conserving flavour oscillations is demonstrated. 

\end{enumerate}

\section{Even-dimensional representations and $\mathcal{T}^2 =-1$}

For $\mathcal{T}^2 =-1$ we have $ZZ^{\ast }=-\mathbb{1}$  where $\mathbb{1}$  is a finite dimensional identity matrix \cite{PhysRevA.82.042101}. For any wave function $\psi$, $ZZ^{\ast }\psi^{\ast } =-\psi^{\ast }.$ If $\psi$ is  an eigenstate of $Z$ with an eigenvalue $\lambda\ne 0$,then $Z^{\ast }\psi^{\ast } =\lambda^{\ast } \psi^{\ast } $. Consequently $$ZZ^{\ast }\psi^{\ast } =\lambda^{\ast } Z\psi^{\ast } =-\psi^{\ast } $$ and so $\psi^{\ast }$ is an eigenstate of $Z$ with eigenvalue $-1/\lambda^{\ast } $. Because of the pairing or doublet structure of eigenstates  it is argued that the matrix $Z$ is even dimensional \cite{PhysRevA.82.042101}.

Following \cite{PhysRevA.82.042101}, we will take the standard approach of defining a \cPT- symmetric quantum theory where the Hamiltonian, $H$,   satisfies the \cPT-axioms:
\begin{enumerate}
\item $\left[ H,\cPT \right] =0$
    \item There is a $\cPT$ inner product with respect to which $H$ is self adjoint.
    \item All the eigenvalues of $H$ are real.
\end{enumerate}
These conditions are $H$-dependent and distinguishes this {\it new } \cPT-symmetric quantum mechanics from the conventional Hermitian one. However, there is no rigorous algorithm to construct a \cPT~inner product. For the Dirac equation, which due to spatial derivatives is not  purely a  matrix model, the proposed inner product \cite{PhysRevA.82.042101} has issues with orthonormality in the Hilbert space.  $\cPT$-inner products  have negative norm states. This defect is corrected (as shown in the \cT-even case) by constructing a related $\cCPT$ inner product, \cb which requires that $\cPT$ is an unbroken symmetry\cbl. This inner product may not be unique \cite{Bender:2009en}. We  construct inner products which are positive definite, lead to completeness relations in the Hilbert space, and thus describe a physical theory.
\vskip .2cm
For a finite dimensional system in the $\cT$- odd case, a basis can be chosen \cite{PhysRevA.82.042101} such that
\be
\label{canonical z}
Z=\left( \begin{matrix}e_{2}&&\\ &\ddots &\\ &&e_{2}\end{matrix}  \right)\quad 
\ee
is an antisymmetric $2N \times 2N$ matrix with $e_{2}=i\sigma_{2} $ \footnote{The Pauli matrices are \, $\sigma_{0} =\left( \begin{matrix}1&0\\ 0&1\end{matrix}  \right) ,\  \   \sigma_{1} =\left( \begin{matrix}0&1\\ 1&0\end{matrix}  \right) ,\  \   \sigma_{2} =\left( \begin{matrix}0&-i\\ i&0\end{matrix}  \right) ,\  {\rm and} \;  \sigma_{3} =\left( \begin{matrix}1&0\\ 0&-1\end{matrix}  \right) $.} . In this basis, parity is a diagonal matrix $S$ of the form \cite{PhysRevA.82.042101}
\be
S=\left( \begin{matrix}\mathbb{1}_{M}&\\ &-\mathbb{1}_{2N-M}\end{matrix}  \right).
\ee
We  consider $M=N$ in applications. For the \cT ~even case, $Z$ is the identity, and the $\cPT$ inner product is  defined as
$$\left( \phi ,\psi  \right)_{\cPT} =\left( \cPT\phi  \right)^{T} \psi =\phi^{\dag } S\psi . $$
In order to maintain this form for the \cT odd case, the modified inner product 
\be
\label{MIP}
\left( \phi ,\psi  \right)_{\cPT} =\left( \cPT\phi  \right)^{T} Z\psi =\phi^{\dag } S\psi 
\ee
is introduced in \cite{PhysRevA.82.042101}. This serves as a guiding principle for the construction of the \cPT-inner product in the Dirac equation. 
Using the \cPT-axioms and Eq.\eqref{MIP}, it can be shown \cite{PhysRevA.82.042101} that the eigenstates of a \cPT-symmetric Hamiltonian, $\mathcal{H}$, with \emph{distinct} eigenvalues are \emph{orthogonal}.  
\vskip .1cm
The $\mathcal{PT}$ self-adjoint condition for the Hamiltonian $\mathcal H$ is 
\begin{equation}
    \begin{aligned}
         \left<{\mathcal H} \phi, \psi\right>_\mathcal{P T}=\left< \phi, \mathcal H \psi\right>_\mathcal{P T},
    \end{aligned}
\end{equation}
 which implies
 \begin{equation}
    \begin{aligned}
         ({\mathcal H}\phi)^\dagger S \psi & = \phi^\dagger S {\mathcal H}\psi, \\
         \phi^\dagger {\mathcal H}^\dagger S \psi &= \phi^\dagger S {\mathcal H}\psi, \\
          {\mathcal H}^\dagger S  &=  S {\mathcal H} ,\\
          \end{aligned}
\end{equation}
and  we obtain the \emph{pseudo-Hermiticity} relation \cite{R16}
\be
\label{Sat}
S^{-1}  {\mathcal H}^\dagger S  =   {\mathcal H}.
\ee
In order to understand the significance of this condition for transition probability amplitudes, consider for any non-Hermitian Hamiltonian $H$, the time evolution operator  
\begin{equation}
    U(t) = e^{-it{ H}}
\end{equation}
acting on a ket. For the standard Dirac inner product the time evolution operator acting on the bra is
\begin{equation}
    U^\dagger(t) = e^{it{ H}^\dagger}
\end{equation}
We find that \footnote{We adopt the convention that the  angular brackets $\left< | \right> $ refer to an inner product for an underlying Dirac Hilbert space.} ,
\begin{equation}
    \langle\phi(t) | \psi(t)\rangle
    = \left\langle\phi(0)\left|e^{i t { H}^{\dagger}} e^{-i t { H}}\right| \psi(0)\right\rangle\neq  \langle\phi(0) | \psi(0)\rangle
\end{equation}
because ${ H} \neq { H}^\dagger$, i.e. $H$ is not self-adjoint with respect to the Dirac inner product. Let $ H^\sharp$ be the adjoint of $H$ in the $\cPT$ inner product \cite{Das:2009it,Das_2011}; then by definition 
\be
\left< H^{\sharp }\phi |\psi  \right>_{\cPT} =\left< \phi |H\psi  \right>_{\cPT}.
\ee
Now 
\be
\left< \phi |H\psi  \right>_{PT} =\left< S^{-1}H^{\dag }S\phi |\psi  \right>_{PT} 
\ee
and so $H^\sharp=S^{-1}H^{\dag }S$. For $H=\mathcal H$ we have from \eqref{Sat} that ${\mathcal H}^\sharp={\mathcal H}$ and so
\be
\label{transition}
\left< \phi \left( t \right) |\psi \left( t \right)  \right>_{PT} =\left< \phi \left( 0 \right) |\psi \left( 0 \right)  \right>_{PT}, 
\ee
which is  a necessary condition for unitary evolution \cite{Das_2011,Das:2009it}. 
This feature is maintained for a   $q$ inner product where $\left< \phi | \psi \right>_q \equiv \left< \phi |q| \psi \right>$ and we  define $H^{\natural }=q^{-1}Hq$. In this inner product, the unitary operator acting on the bra is $ U^\natural(t) \equiv e^{itH^\natural}$.
The amplitude  $\langle\phi(t) | \psi(t)\rangle_q$ satisfies the equivalent of \eqref{transition}. Because the $\cPT$ inner product has states of negative norm, such a new $q$ inner product is used when considering transition probabilities and is given below. 

\vskip .2cm

A physical $q$ inner product is the $\cCPT$ inner product, which is defined for \cT-even \cite{Bender:2004guu}  as $\left( \phi ,\psi  \right)_{\cCPT} =\left( \cCPT\phi  \right)^{T} .\psi $, where $\cC$ satisfies $\left[ \cC,\mathcal H \right] =0,\  \  \left[ \cC,\cPT \right] =0$ and $\cC^{2}=1$. For finite dimensional Hamiltonians and the \cT-odd case, it is \emph{modified}~\cite{PhysRevA.82.042101}  to  $$\left( \phi ,\psi  \right)_{\cCPT} =\left( \cCPT\phi  \right)^{T}Z \psi .$$ Observables are defined to be self-adjoint with respect to the $\cCPT$ inner product.
Based on the above \cPT-axioms for the $\cT$-odd case, the generic \cPT-invariant Hamiltonian has the form~\cite{PhysRevA.82.042101}
\be
{\mathcal H}=\left( \begin{matrix}A&iB\\ iB^{\dag }&D\end{matrix}  \right) ,
\ee
where $A$, $D$ are square Hermitian matrices, and $B$ is a real quaternion \cite{Stillwell2008-rx}. We will consider in detail the Hilbert space structure and unitarity for the Hamiltonian with $A=-D=a_{0}\sigma_{0}$ and $a_{0}$ is real \cite{PhysRevA.82.042101}. We denote this  4-dimensional simplified model as SFDM. Such a model allows the study of unconventional flavour oscillations for non-relativistic systems \cite{Bittencourt2021-ea}.

 \section{$\cT$ odd relativistic quantum mechanics}
 \cb We  consider the generalisation of the considerations in the previous sections to $\cPT$-symmetric relativistic fermionic systems. \cbl Spin {1/2} fermions satisfy the Dirac equation, which, however, is no longer a simple matrix system because of the presence of spatial derivatives. The $\cPT$ axioms  in the last section have been applied to the Dirac equation in \cite{Jones-Smith:2009qeu}. However, because of the presence of spatial derivatives in the Hamiltonian, it is possible to consider different $\cPT$-inner products. We present a new $\cPT$ inner product, which allows construction of a non-Hermitian Dirac equation identical to the one in \cite{Jones-Smith:2009qeu}, but leads to a different  orthogonality relation in the Hilbert space. We  compare our inner product with that in \cite{Jones-Smith:2009qeu}.

\cm
\subsection{The $\mathcal{PT}$ -Symmetric Dirac Hamiltonian}

The Dirac Hamiltonian\footnote{In terms of the usual Dirac gamma matrices $\gamma^{i} =\beta \alpha^{i}$ and $\gamma^{0} =\beta $.} is given by 
\begin{equation}
\label{Dirac Hamiltonian sec 2}
    H_D=-i \boldsymbol{\alpha} \cdot \nabla+\beta,
\end{equation}
where the $4 \times 4$ matrices $\boldsymbol \alpha$ and $\beta$ satisfy the Clifford algebra \footnote{The non-trivial Hermitian irreducible representation of $\boldsymbol \alpha$ and $\beta$ is through $4\times4$ matrices \cite{Das2014-tf} in four spacetime dimensions. The Clifford algebra for Dirac gamma matrices is $\left\{ \gamma^{\mu } ,\gamma^{\nu }  \right\} =2\eta^{\mu \nu } $ where $\eta^{\mu \nu }$ is the flat space metric}. We start by reviewing a non-Hermitian but $\cPT$-symmetric version of $ H_D$ found  in the context of a particular  $\cPT$ inner product \cite{Jones-Smith:2009qeu}. We consider a different inner product but follow similar steps. Hence, we  briefly outline the common steps, referring to \cite{Jones-Smith:2009qeu} along the way. Using our inner product we discuss the unitarity of the theory and the methods used to maintain the covariance of $H_D$ under the Poincar\'e group \footnote{ The \cPT operator has an interesting property in the context of the complex Lorentz group, which is crucial in proving the spin-statistics theorem \cite{Streater:1989vi}. The real Lorentz group $O\left( 3,1 \right) $ has four topologically disconnected pieces \cite{R14a}. The $\cPT$ operator does not belong to the piece corresponding to the proper orthochronous real Lorentz group. For the complex Lorentz group this is no longer the case \cite{R2}. } .
 In \cite{PhysRevA.82.042101} a general characterisation of four dimensional {\it $\mathcal{PT}$-symmetric matrices} is given but  $H_D$ involves differential operators, which are treated by assuming  that the solution of the Dirac equation \cb ($i\frac{\partial \Psi }{\partial t} =H_{D}\Psi $) has the form

\begin{equation}
\label{Dirac}
    \Psi=\left(\begin{array}{l}
\psi_1 \\
\psi_2 \\
\psi_3 \\
\psi_4
\end{array}\right)
e^{-i(E-\mathbf{p.x})}
\equiv \psi(\bf p) e^{-i(E-\mathbf{p.x})}.
\end{equation}
\cbl where $\psi(\bf p)$ is the Dirac spinor. For such a solution $H_{D}= \boldsymbol{\alpha} \cdot \mathbf{p}+\beta m$
where $\mathbf{p} = (p_x, p_y, p_z)$ is the 3-component momentum of the particle. Following \cite{PhysRevA.82.042101} we represent the  parity as ${\cP} = S$ and time-reversal as ${\cT} = Z \hat{K}$, 
and $S$ and $Z$ are \emph{linear} operators. We will summarise  the framework of the reducible $\cPT$ symmetric representation \cite{Jones-Smith:2009qeu,PhysRevA.82.042101} of the Clifford algebra \cite{Das2014-tf} before we  consider the novel aspects of $\cPT $-symmetric inner products.

\subsection{General Conditions}
It is known that for spinors \cite{R14a},  the generator of Lorentz boosts ($K_i$) and rotations ($J_i$) (where $i$ represents$(x,y,z)$) can be expressed in terms of the Dirac gamma matrices as 
\begin{equation}
\label{K to a}
    K_i \rightarrow \frac{i}{2} \alpha_i, \quad J_x \rightarrow-\frac{i}{2} \alpha_y \alpha_z, \quad J_y \rightarrow-\frac{i}{2} \alpha_z \alpha_x, \quad J_z \rightarrow-\frac{i}{2} \alpha_x \alpha_y.
\end{equation}
Under the action of $\cP$ and $\cT$ we expect 
\begin{equation}
    \begin{aligned}
    \label{commutation relation between P and K}
        \hat{\cP}^{-1} K_i \hat{\cP} &= - K_i, \\
        \hat{\cT}^{-1} K_i \hat{\cT} &= - K_i.
    \end{aligned}
\end{equation}
Because $\hat{\cP}^2 = 1$ we have $\hat{\cP}^{-1} = \hat{\cP} = S$. However, since $\hat{\cT}^2 = Z \overline{Z} = -1$, we have $Z^{-1} = - \overline{Z}$, and  $\hat{\cT}^{-1}\psi = - Z \psi$. This is consistent since $\hat{\cT}^{-1}\hat{\cT} \psi=\hat{\cT}^{-1} Z \overline{\psi}= -Z \overline{Z} \psi = \psi $. On using the relations  Eq. (\ref{K to a}) and Eq. (\ref{commutation relation between P and K}), the following holds: $S \alpha_i S = - \alpha_i$ and $-Z \overline{\alpha}_i \overline{Z} = - \alpha_i$, which can be written in a more convenient form as
    \begin{align}
        \label{sa=-as}
        S \alpha_i &= - \alpha_i S, \\
        \label{za=-az}
        Z \overline{\alpha}_i  &= - \alpha_i Z.
    \end{align}

Moreover $[\cP,\cT] = 0$ implies $SZ = Z\overline{S}$. 
On requiring $[H_D,\cPT]=0$, we have $H_D S Z = S Z \overline{H}_D$ and so \cite{Jones-Smith:2009qeu} 
\begin{equation}
\label{H PT commute}
\begin{array}{r}
\alpha_i S Z=S Z \overline{\alpha}_i ,\\
\beta S Z=S Z \overline{\beta}.
\end{array}
\end{equation}

\subsubsection{$\cPT$ self-adjointness  of the Hamiltonian}

We  now impose on $H_D$ the condition that it is $\cPT$ self-adjoint, one of the  axioms of $\cPT$-symmetric quantum mechanics.
The $\cPT$-self-adjoint condition is 
\begin{equation}
\label{PT self adjoint condition}
    (H_{D}\, u(\mathbf{k}) \exp (i \mathbf{k} \cdot \mathbf{r}), v(\mathbf{p}) \exp (i \mathbf{p} \cdot \mathbf{r}))_{\cPT}=(u(\mathbf{k}) \exp (i \mathbf{k} \cdot \mathbf{r}), H_{D}\, v(\mathbf{p}) \exp (i \mathbf{p} \cdot \mathbf{r}))_{\cPT}.
\end{equation}
 We define our $\cPT$ inner product (CS) and compare it with the inner product of Jones-Smith and Mathur (JSM) \cite{Jones-Smith:2009qeu}:
\begin{equation}
    \begin{aligned}
        \text{JSM:} &= \left< \phi \mid \psi  \right>_{\cPT}  =\  (\mathcal{PT}\phi )^{T} Z  \psi \xlongequal{\textcolor{black}{spinor}\, part} \  \phi^\dagger S^\dagger \psi  \\
        \text{CS:} &= \left( \phi \mid \psi  \right)_{\cPT}  =\  (\mathcal{P} \phi )^{\dagger}  \psi \xlongequal{\textcolor{black}{spinor}\,part} \  \phi^\dagger S^\dagger \psi  
        \end{aligned}
\end{equation}
For theories where $\psi$ and $\phi$  are \emph{independent} of momentum $\mathbf{p}$, the final expression for the two inner products is \emph{identical}. However, for the Dirac case, the eigenstates are functions also of the momentum $\mathbf{p}$, and so we need to consider the exponential (plane wave) part in the inner product. Consider \cb states
\begin{equation*}
\label{eigenstates of Dirac Hamiltonian}
    \Psi(\mathbf{k},\mathbf{r})= \psi(\mathbf{k}) e^{i \mathbf{k}\cdot \mathbf{r}} \;{\rm{and}}\; \Phi(\mathbf{p},\mathbf{r})= \phi(\mathbf{p})e^{i \mathbf{p}\cdot \mathbf{r}}. 
    \end{equation*}
    \cbl
Under the $\cPT$ transformation $\mathbf{p}$ is invariant, but  $\mathbf{p}$ picks up a minus sign under $\cP$ . Hence, for the JSM inner product we have
\begin{equation}
    \begin{aligned}
        \,\left< \Phi \mid \Psi  \right>_{\cPT}&=\  (\cPT\Phi )^{T} Z  \Psi \\
           &=(\cPT \psi(\mathbf{k}) e^{i \mathbf{k}\cdot \mathbf{r}} )^{T} Z  \phi(\mathbf{p})e^{i \mathbf{p}\cdot \mathbf{r}} \\
&=\int d \mathbf{r} e^{i \mathbf{k}\cdot \mathbf{r}}e^{i \mathbf{p}\cdot \mathbf{r}}\phi(\mathbf{k})^\dagger S^\dagger \psi(\mathbf{p})\\
&=\delta(\mathbf{k} +\mathbf{p})\phi(\mathbf{k})^\dagger S^\dagger \psi(\mathbf{p}) .
\end{aligned}
\end{equation}
For the CS inner product
\begin{equation}
    \begin{aligned}
         \,\left( \Phi \mid \Psi  \right)_{\cPT}&=\  (\cP\Phi )^{\dagger}  \Psi \\
           &=\  (\cP \psi(\mathbf{k}) e^{i \mathbf{k}\cdot \mathbf{r}} )^{\dagger}  \phi(\mathbf{p})e^{i \mathbf{p}\cdot \mathbf{r}} \\
&=\int d \mathbf{r} e^{-i \mathbf{k}\cdot \mathbf{r}}e^{i \mathbf{p}\cdot \mathbf{r}}\phi(\mathbf{-k})^\dagger S^\dagger \psi(\mathbf{p})\\
&=\delta(\mathbf{k} -\mathbf{p})\phi(\mathbf{-k})^\dagger S^\dagger \psi(\mathbf{p}). 
\end{aligned}
\end{equation}
\vskip .1cm
The JSM inner product is nonvanishing for  $\mathbf{k} = -\mathbf{p}$, 

and for the CS inner product, we have $\mathbf{k} = \mathbf{p}$, which gives $\left( \Phi \mid \Psi  \right)_{PT}=\phi(\mathbf{-k})^\dagger S^\dagger \psi(\mathbf{p}) $ as well. Clearly, the JSM inner product of a state with momentum $\mathbf{p}$ with itself vanishes, which is not physical. The CS inner product of a state with momentum $\mathbf{p}$ with itself is nonvanishing. We should note that for a theory with $\mathbf{p}$-independent states, the JSM inner product and the CS inner product coincide. Moreover $Z$ does not appear explicitly in either inner product.
\cb
\vskip .2cm
In Appendix \ref{sec.appendix self adjointness}, we show that both the JSM and CS inner products lead to the same relations
\begin{align}
   -  \boldsymbol{\alpha}^\dagger  S^\dagger  &=S^\dagger \boldsymbol{\alpha}\\
 \beta^\dagger S^\dagger  &=   S^\dagger \beta ,
\end{align}
 on requiring $H_D$ to be self-adjoint. Using \eqref{H PT commute} we can rewrite these equations \cite{Jones-Smith:2009qeu} as
 \begin{align*}
    \beta  &=- Z^T \beta^T Z^\dagger ,\\
    \alpha_i  &= Z^T \alpha_i^T Z^\dagger.
 \end{align*}
 
 \cbl \noindent We note that \emph{no} requirement of Hermiticity for the gamma matrices is imposed. On comparing  \eqref{sa=-as} with $ -  \boldsymbol{\alpha}^\dagger  S^\dagger  =   S^\dagger \boldsymbol{\alpha} $, we have $\alpha_i = \alpha_i^\dagger$ and so the $\alpha_i$ are always  Hermitian, but the form of $\beta$  distinguishes between a \cPT- symmetric and a Hermitian theory in an 8-dimensional representation (for the $\gamma$ matrices) of the Clifford algebra .

\section{Hilbert structure of $\cCPT$ inner products}

In order for Hamiltonians, which satisfy the $\cPT$ axioms, to be physically significant, we need to check that a Hilbert space structure  arises. For a Hilbert space $\mathfrak H$, we require an  inner product $(,)$  such that  any vectors (states) $x$, $y$ and $z$  satisfy the  conditions
\begin{enumerate}
    \item $\left( x,x \right) >0$.
      \item $\left( x,y \right) =\overline{\left( y,x \right) } $.
  \item $\left( z,ax+by \right) =a\left( z,x \right) +b\left( z,y \right)$
     where $a,b \in \mathbb{C}$.
\end{enumerate}
 Such an inner product leads to  an orthonormal basis,  a completeness relation and a physical theory which conserves probability. This is investigated and demonstrated for the two Hamiltonians discussed above :
\begin{itemize}
\item[] (a) the matrix model SFDM, which is $\mathbf{p}$-independent
\item[] and
\item[] (b) the restricted 8-dimensional \cPT-symmetric Dirac Hamiltonian (${\rm Dirac\,8D}$), which is $\mathbf{p}$-dependent.
\end{itemize}
The inner products constructed will be  $\cCPT$ inner products (derived from  $\cPT$ inner products), and will be shown to satisfy the above Hilbert space criteria. The \emph{existence} of physical inner products is also a necessary condition for constructing $\cPT$-symmetric field theories using path integrals \cite{R3a,Croney:2023gwy}.

\subsection{Hilbert space for SFDM}

The $\cPT$-invariant four dimensional Hamiltonian $H_C$ in the canonical basis \cite{PhysRevA.82.042101} for SFDM is
\vskip .1cm
\be
\label{SFDM}
H_{C}=\left( \begin{matrix}a_{0}\sigma_{0} &ib\\ ib^{\dag }&-a_{0}\sigma_{0} \end{matrix}  \right), 
\ee
\vskip .1cm
\noindent where $b=b_{0}\sigma_{0} +ib_{1}\sigma_{1} +ib_{2}\sigma_{2} +ib_{3}\sigma_{3} $ {\rm and} $a_{0},b_{0},b_{1},b_{2},{\rm and}\,  b_{3}$ are real. This is the simplest non-trivial $\cT$ odd  $\cPT$-symmetric Hamiltonian.  The $\cP$ and $\cT$ operators are \footnote{We write $\mathbb{1}_{n}$ for a $n$-dimensional identity matrix. } 
 \be
 \cP=\left( \begin{matrix}\mathbb{1}_{2}&0\\ 0&-\mathbb{1}_{2}\end{matrix}  \right) \  {\rm and}\  \  \cT=\left( \begin{matrix}e_{2}&0\\ 0&e_{2}\end{matrix}  \right).  
 \ee
 \vskip .1cm
  \noindent The eigenvalues $\lambda_{i}$ and eigenvectors $|e_{i})$ of $H_{C}$ satisfy
\be
H_{C}|e_{i})=\lambda_{i} |e_{i}),\quad i=1,\ldots,4.
\ee
We consider $\det H_{C}=1$. \footnote{A useful parametrisation \cite{PhysRevA.82.042101} of $a_0$ and $b_\mu$ , $\mu=0,1,2,3$ is
 \begin{align*}
   b_{0}&=\sinh \chi \cos \psi, \\ 
   b_{1}&=\sinh \chi \sin \psi \sin \theta \cos \phi, \\
   b_{2}&=\sinh \chi \sin \psi \sin \theta \sin \phi,\\
   b_{3}&=\sinh \chi \sin \psi \cos \theta ,\\
   a_{0}&=\cosh \chi.
 \end{align*}}
 The eigenvalues are $\lambda_{1}= \lambda_{2}=-1,{\rm and} \; \lambda_{3}= \lambda_{4}=1 $. The corresponding eigenvectors are given in Appendix \ref{HCeigen}.

 \subsubsection{Construction of the inner product}

 We define the $\cPT$ bra as
 \be
 \label{SFbra}
 \left( \phi  \right] \equiv\left( \cPT|\phi ) \right)^{T} Z.
 \ee
The resulting \cPT~inner product is:
\be
\left( e_{i}|e_{j} \right) =\begin{cases}
 -\delta_{ij} ,&\text{if $i,j=1,2$ }\\ 
 \delta_{ij},&\text{if $i,j=3,4$ }
\end{cases},
\ee
which is not positive definite.
Following the procedure used for even $\cT$ \cite{Bender:2003fi,Bender:2003gu,Weigert:2003py}, we introduce a $\cC$ operator 
\be
\cC=\sum^{4}_{j=1} |e_{j})(e_{j}|.
\ee
where the bra is the $\cPT$ adjoint of a ket.
The next step is to introduce a $\cCPT$ inner product \cite{PhysRevA.82.042101} denoted by $\left(( | )\right) $:
\be
\left(( \phi |\psi  )\right) \equiv \left( \cCPT\phi  \right)^{T} Z\psi ,
\ee
where $\left( \cCPT\phi  \right)^{T} Z$ is the $\cCPT$ bra $(( \phi |$. It is easy to check that 
\be
\left(( e_{i},e_{j} )\right) = \delta_{ij}
\ee
(which is positive definite) and that the Hilbert space conditions hold. We have the completeness condition on the Hilbert space
\be
\sum^{4}_{j=1} |e_{j}))((e_{j}|=\mathbb{1},
\ee
where $|e_{j}))$ is identified with $|e_{j})$.
\vskip .2cm
 ~In the language of flavour oscillation models, the $|e_{j})$ are the analogue mass eigenstates  and the coordinate basis vectors (such as $\left( 1,0,0,0 \right)^{T}$) are, naively, the flavour states. We could write such flavour eigenstates as
\begin{equation}
\begin{aligned}
    f_1 = \left(\begin{array}{c}
1 \\
0 \\
 0 \\
0 \\
\end{array}\right) ,\quad 
 f_2 = \left(\begin{array}{c}
0  \\
1 \\
0 \\
0 \\
\end{array}\right) ,\quad
    f_3 = \left(\begin{array}{c}
0 \\
0 \\
1\\
0 \\
\end{array}\right) ,\quad {\rm and} \quad
 f_4 = \left(\begin{array}{c}
0 \\
0 \\
0 \\
1 \\
\end{array}\right),
\end{aligned}
\end{equation}
following the usual procedure used in the study of flavour in Hermitian theories. \cb A related choice was used in \cite{Ohlsson:2019noy}.
 The ``mass'' eigenstates can be expressed as  linear superposition of the naive flavour eigenstates, i.e.
\begin{equation}
    \left| e_l \right)) = \sum_j\alpha_{lj}  \left| f_j \right)).
\end{equation}

The completeness relation becomes
\begin{equation}
    \sum_{j,k} B_{jk}  \left| f_j \right))\left(( f_k \right| = \mathbb{1}.
\end{equation}
where $B_{jk} \equiv \sum_l \alpha_{lj} \overline{\alpha}_{lk}$.
Clearly the naive flavour eigenstates do not, in general, satisfy the canonical completeness relation under the $\cCPT$ inner product (unless $B_{jk}=\delta_{jk}$) and so probability conservation in oscillations of such ``flavour'' states would not be manifest.
\vskip .2cm
 We  construct alternative flavour eigenstates such as 

\begin{equation}
    \begin{aligned}
        \left|f'_{1}\right)) & = \cos \theta\left|e_1\right)) + \sin \theta\left|e_3\right)) ,\\
                \left|f'_2\right)) & = \cos \theta \left|e_2\right)) + \sin \theta \left|e_4\right)), \\
                \left|f'_3\right)) & =  -\sin \theta\left|e_1\right)) +\cos \theta \left|e_3\right)), \\
                \left|f'_4\right)) & = -\sin \theta\left|e_2\right)) +\cos \theta \left|e_4\right)) ,\\
    \end{aligned}
\end{equation}
where $\theta$ being a mixing angle.
Since the $\left|e_i\right))$ are orthogonal to each other, it is trivial to see, for example, that $\left|f'_{1}\right))$ should to orthogonal to $\left|f'_{2}\right))$ and  $\left|f'_{4}\right))$, and the $\left|f'_{i}\right))$ obey a completeness relation.

So they can be treated as suitable flavour eigenstates, which oscillate among themselves. These states show flavour oscillations that conserve probability. The orthogonality and completeness relations resolve the issue of conservation of probabilities in mixing phenomena for $\cPT$- symmetric theories \cite{Ohlsson:2019noy}. 
\vskip .5cm
\noindent Next we  look at the time evolution of the flavour states

\begin{equation}
    \begin{aligned}
        \left|f'_{1}(t) \right)) & = e^{-i\lambda_1 t}\cos \theta\left|e_1\right)) + e^{-i\lambda_3 t}\sin \theta\left|e_3\right)), \\
                \left|f'_2(t)\right)) & =e^{-i\lambda_2 t}\cos \theta\left|e_2\right)) + e^{-i\lambda_4 t} \sin \theta\left|e_4\right)), \\
                \left|f'_3(t)\right)) & = -e^{-i\lambda_1 t}\sin \theta\left|e_1\right)) + e^{-i\lambda_3 t}\cos \theta\left|e_3\right)), \\
                \left|f'_4(t)\right)) & = -e^{-i\lambda_2 t}\sin \theta\left|e_2\right)) + e^{-i\lambda_4 t}\cos \theta\left|e_4\right)), \\
    \end{aligned}
\end{equation}
\noindent where $\lambda_i$ are the eigenvalues, and we  calculate the transition probability using the usual definition $P(f_i \rightarrow f_j) = |(\left(f'_j|f'_i(t)\right))|^2$. The result is shown in table \ref{transition probability SFDM}.

\vskip 1cm

\begin{table}[htbp]
    \centering
    \begin{tabular}{|c|c|c|c|c|} \hline  
         $P$&  $\left|f'_{1}(t)\right))$&  $\left|f'_{2}(t)\right))$&  $\left|f'_{3}(t)\right))$& $\left|f'_{4}(t)\right))$\\ \hline  
         $\left|f'_{1}\right))$&  $\cos ^2(2 \theta ) \sin ^2(t)+\cos ^2(t)$&  0&  $\sin ^2(2 \theta ) \sin ^2(t)$& 0\\ \hline  
         $\left|f'_{2}\right))$&  0&  $\cos ^2(2 \theta ) \sin ^2(t)+\cos ^2(t)$&  0& $\sin ^2(2 \theta ) \sin ^2(t)$\\ \hline  
         $\left|f'_{3}\right))$&  $\sin ^2(2 \theta ) \sin ^2(t)$&  0&  $\cos ^2(2 \theta ) \sin ^2(t)+\cos ^2(t)$& 0\\ \hline  
         $\left|f'_{4}\right))$&  0&  $\sin ^2(2 \theta ) \sin ^2(t)$&  0& $\cos ^2(2 \theta ) \sin ^2(t)+\cos ^2(t)$\\ \hline 
    \end{tabular}
    \caption{transition probability of the flavour states for SFDM}
    \label{transition probability SFDM}
\end{table}

\noindent We note that the sum of the transition probabilities of any state $\left|f'_{i}\right))$ to all other flavour states is $1$ . 

\cbl

\subsection{Hilbert space for Dirac 8D}
The \cPT-symmetric Model 8 of \cite{Jones-Smith:2009qeu} has the Hamiltonian $H_{8}$ in the momentum basis:
\vskip .1cm
\be
H_{8}=\left( \begin{matrix}\mathbf{\sigma .p}&0&\left( m_{0}+m_{3} \right) \sigma_{0} &\left( m_{1}-im_{2} \right) \sigma_{0} \\ 0&\mathbf{\sigma .p}&\left( m_{1}+im_{2} \right) \sigma_{0} &\left( m_{0}-m_{3} \right) \sigma_{0} \\ \left( m_{0}+m_{3} \right) \sigma_{0} &\left( m_{1}+im_{2} \right) \sigma_{0} &-\mathbf{\sigma .p}&0\\ \left( m_{1}-im_{2} \right) \sigma_{0} &\left( m_{0}-m_{3} \right) \sigma_{0} &0&-\mathbf{\sigma .p}\end{matrix}  \right),
\ee
\vskip .2cm
\noindent where $m_0,m_1,m_2$ and $m_3$ are real parameters. For simplicity we concentrate on two non-Hermitian Hamiltonians derived from $H_8$  on restricting  parameters: $H_{8r}=H_{8}|_{m_{3}=0}$ and  $H_{8v}=H_{8}|_{m_{1}=m_{3}=0}$.
For both cases we show that it is possible to have an inner product on the Hilbert space (satisfying the Hilbert space axioms). The analysis of cases with $\mathbf{p}=0$ is similar to that for $H_C$. The case $\mathbf{p} \neq 0$ will be considered for $H_{8v}$ (for simplicity). \cm Following \cite{Jones-Smith:2009qeu,JonesSmith2010a}, on using the Dirac representation of the gamma matrices, we have for $S$ 
\vskip .1cm
\begin{equation}
       S = \left(
\begin{array}{cccc}
 0 & 0 & \mathbb{1} & 0 \\
 0 & 0 & 0 & \mathbb{1} \\
 \mathbb{1} & 0 & 0 & 0 \\
 0 & \mathbb{1} & 0 & 0 \\
\end{array}
\right).
\end{equation}

\cbl
\subsubsection{Model $H_{8v}$, $\mathbf{p} =0$}
\noindent The Hamiltonian $H_{8v}$ for $\mathbf{p} =0$ is:
\begin{equation}
    H = \left(
\begin{array}{cccc}
 0 & 0 & m_0 & -i m_2 \\
 0 & 0 & i m_2 & m_0 \\
 m_0 & i m_2 & 0 & 0 \\
 -i m_2 & m_0 & 0 & 0 \\
\end{array}
\right)
\end{equation}
where each entry is muliplied by $\mathbb{1}_2$.
\vskip .2cm
\noindent The eigenvectors of the Hamiltonian (for $m_{0}>m_{2}$) are
\begin{equation}
\begin{aligned}
    v_1 = \left(\begin{array}{c}
\frac{i m_2}{b^2} \\
-\frac{m_0}{b^2} \\
 0 \\
1 \\
\end{array}\right) ,\quad 
 v'_2 = \left(\begin{array}{c}
-\frac{m_0}{b^2} \\
-\frac{i m_2 }{b^2} \\
1 \\
0 \\
\end{array}\right), \\
    u_1 = \left(\begin{array}{c}
-\frac{i m_2}{b^2} \\
\frac{m_0}{b^2} \\
 0 \\
1 \\
\end{array}\right) ,\quad 
 u'_2 = \left(\begin{array}{c}
\frac{m_0}{b^2} \\
\frac{i m_2 }{b_0^2} \\
1 \\
0 \\
\end{array}\right),
\end{aligned}
\end{equation}
where $u_1$ and $u'_2$ correspond to the eigenvalue $b^2=\sqrt{m_0^2-m_2^2}$, while $v_1$ and $v'_2$ correspond to the eigenvalue $-\sqrt{m_0^2-m_2^2}$. As for the case $H_C$, the eigenvectors are degenerate and so any linear combination of the eigenvectors with the same eigenvalue would still be an eigenvector.  We define $u_2$ and $v_2$: 
\begin{align}
    u_2 &= u_1 + \frac{i m_0}{m_2} u'_2 , \\
    v_2 &= v_1 + \frac{i m_0}{m_2} v'_2   
    \end{align}
 so that $u_1, u_2, v_1$ and $ v_2$ are mutually orthogonal  under the $\mathcal{PT}$ inner product. Since the $\mathcal{PT}$ inner product can be negative, the generic normalised ket is defined as $|e_{n}\rangle  \equiv \frac{|e\rangle}{\sqrt{|\left<e|e\right>|}}$ to ensure the argument inside the square root to be positive.

 \vskip .2cm
 In terms of the kets of these normalised eigenvectors, the $\cC$ operator can again be defined  by
 \begin{equation}
\label{C operator bender h8v}
\begin{aligned}
    \mathcal{C} &= |u_{1n}\left)\right( u_{1n}| + |u_{2n}\left)\right( u_{2n}| + |v_{1n}\left)\right( v_{1n}|_\mathcal{PT}  + |v_{2n}\left)\right( v_{2n}|   \\
    &= (|u_{1n}\left>\right<\mathcal{PT} u_{1n}| Z) + (|u_{2n}\left>\right<\mathcal{PT} u_{2n}| Z) + (|v_{1n}\left>\right<\mathcal{PT} v_{1n}| Z) + (|v_{2n}\left>\right<\mathcal{PT} v_{2n}| Z)\\
    &=
    \frac{1}{\sqrt{m_0^2-m_2^2}} \left(
\begin{array}{cccc}
 0 & 0 & m_0 & -i m_2 \\
 0 & 0 & i m_2 & m_0 \\
 m_0 & i m_2 & 0 & 0 \\
 -i m_2 & m_0 & 0 & 0 \\
\end{array}
\right)
\end{aligned}
\end{equation}
 and the associated  $\mathcal{CPT}$ inner product is  $((\phi, \psi)) \equiv(\mathcal{C P} \mathcal{T} \phi)^{\mathcal{T}} Z\psi $. The completeness relation is
\begin{equation}
    |u_{1n}\left))(\right( u_{1n}| + |u_{2n}\left))(\right( u_{2n}| + |v_{1n}\left))(\right( v_{1n}|  + |v_{2n}\left))(\right( v_{2n}|={\mathbb I}.
\end{equation}
and has the same implications for conservation of probabilities in  flavour oscillations discussed earlier.

\subsubsection{Model $H_{8r}$, $\mathbf{p} =0$}
The Hamiltonian $H_{8r}$ for $\mathbf{p} =0$ 
 \begin{equation}
    H = \left(
\begin{array}{cccc}
 0 & 0 & m_0 & m_1-i m_2 \\
 0 & 0 & m_1 + i m_2 & m_0 \\
 m_0 & m_1 + i m_2 & 0 & 0 \\
 m_1-i m_2 & m_0 & 0 & 0 \\
\end{array}
\right).
\end{equation}
The eigenvalues $\{\lambda_{i},i=1,\ldots,4\}$ for $H$ are real (for $m_0 >m_2$ ) and are not degenerate:
\begin{align}
   \lambda_1 &=-\lambda_4=-m_1-\sqrt{m_0^2-m_2^2},\nonumber\\  \lambda_2 &=-\lambda_3 =-\sqrt{m_0^2-m_2^2}+m_1 ,\nonumber\\   
\end{align}
where the corresponding normalised eigenvectors are $\tilde v_1, \tilde v_2, \tilde u_1, \tilde u_2$:
\begin{equation}
\begin{aligned}
   \tilde v_1 = n\left(\begin{array}{c}
-1 \\
-a \\
-\tfrac{c+d}{m_0 \lambda{_1}} \\
1 \\
\end{array}\right) ,\quad 
 \tilde v_2 = n\left(\begin{array}{c}
 1 \\
-\overline{a} \\
-\tfrac{d-c}{m_0 \lambda_2} \\
1 \\
\end{array}\right), \\ \quad
    \tilde u_1 = n\left(\begin{array}{c}
-1 \\
\overline{a} \\
-\tfrac{d-c}{m_0 \lambda_2} \\
1 \\
\end{array}\right) ,\quad 
 \tilde u_2 =n\left(\begin{array}{c}
 1 \\
a \\
-\tfrac{c+d}{m_0 \lambda_1} \\
1 \\
\end{array}\right)
\end{aligned}
\end{equation}
with $a=\left( \sqrt{m^{2}_{0}-m^{2}_{2}} +im_{2} \right) /m_{0}$,\;\,$\overline{a}=( \sqrt{m^{2}_{0}-m^{2}_{2}} -im_{2} ) /m_{0}$,\;$c=(m^{2}_{0}-m^{2}_{2}+im_{1}m_{2})$ \;  $d=\left( m_{1}+im_{2} \right) \sqrt{m^{2}_{0}-m^{2}_{2}} $, and $n=\frac{1}{2 (1-\frac{m_2^2}{m_0^2})^\frac{1}{4}} $.
The  $\cC$ operator,  constructed as in \eqref{C operator bender h8v}, is
\begin{equation}
\label{C operator bender h8r}
\begin{aligned}
     \widetilde {\mathcal{C}} 
    &=
    \frac{1}{\sqrt{m_0^2-m_2^2}} \left(
\begin{array}{cccc}
 0 & 0 & m_0 & -i m_2 \\
 0 & 0 & i m_2 & m_0 \\
 m_0 & i m_2 & 0 & 0 \\
 -i m_2 & m_0 & 0 & 0 \\
\end{array}
\right).
\end{aligned}
\end{equation}
Furthermore unitarity or conservation of probability needs to be maintained for the theory to be physical. We have seen that the $\cCPT$ inner product has a completeness relation necessary for unitarity.

\subsubsection{Model $H_{8v}$, $\mathbf{p} \neq 0$}
The Hamiltonian $H_{8v}$ for  Model 8 with  restricted parameters and $\mathbf{p} \neq0$ provides a test of the usefulness of the  \cm CS  \cbl inner product. 
\begin{equation}
    H_{8v} = \left(
\begin{array}{cccc}
 \mathbf{\sigma.p} & 0 & m_0 & -i m_2 \\
 0 & \mathbf{\sigma.p} & i m_2 & m_0 \\
 m_0 & i m_2 & -\mathbf{\sigma.p} & 0 \\
 -i m_2 & m_0 & 0 & -\mathbf{\sigma.p} \\
\end{array}
\right).
\end{equation}
\cm  For simplicity, we  divide the eigenvectors as a direct product of two subspaces: the first subspace is formed from the eigenvectors of the helicity operator $\mathbf{\sigma.\hat{n}}$:
\begin{align}
    \xi_{+}(\hat{\mathbf{n}})=\left(\begin{array}{c}
\cos \theta / 2 \\
\exp (i \varphi) \sin \theta / 2
\end{array}\right), \quad \xi_{-}(\hat{\mathbf{n}})=\left(\begin{array}{c}
-\exp (-i \varphi) \sin \theta / 2 \\
\cos \theta / 2
\end{array}\right),
\end{align}
where the direction of motion $\hat{n}$ is parameterised by
\begin{equation}
    \begin{aligned}
& \hat{\mathbf{n}}_x=\sin \theta \cos \varphi, \\
& \hat{\mathbf{n}}_y=\sin \theta \sin \varphi, \\
& \hat{\mathbf{n}}_z=\cos \theta .
\end{aligned}
\end{equation}
Hence we halve the effective dimension of the matrix, and the second subspace is formed by  the eigenvectors of the $4 \times 4$ Hamiltonian
\begin{equation}
    H_{8v} = \left(
\begin{array}{cccc}
 p & 0 & m_0 & -i m_2 \\
 0 & p & i m_2 & m_0 \\
 m_0 & i m_2 & -p & 0 \\
 -i m_2 & m_0 & 0 & -p \\
\end{array}
\right),
\end{equation}
\cbl
whose normalised eigenvectors can be chosen as

\begin{equation}
\begin{aligned}
    \widehat{v}_1 = \frac{1}{\sqrt{2m_{0}\epsilon}} \left(\begin{array}{c}
\frac{i m_2 \left(\epsilon-p\right)}{m_{\text{eff}}} \\
\frac{m_{0}\left( p-\epsilon \right) }{m_{\text{eff}}}  \\
 0 \\
m_{\text{eff}} \\
\end{array}\right) ,\quad 
 \widehat{v}_{2} = \frac{1}{\sqrt{2m_{0}\epsilon} }\left(\begin{array}{c}
 i\left( p-\epsilon \right)   \\
0 \\
im_0 \\
m_2 \\
\end{array}\right) \\
   \widehat{u}_{1} =\frac{1}{\sqrt{2m_{0}\epsilon}} \left(\begin{array}{c}
 -\frac{im_{2}\left( \epsilon+p \right) }{m_{\text{eff}}}  \\
\frac{m_{0}\left( p+\epsilon \right) }{m_{\text{eff}}} \\
 0 \\
m_{\text{eff}} \\
\end{array}\right) ,\quad 
 \widehat{u}_{2} = \frac{1}{\sqrt{2m_{0}\epsilon} }\left(\begin{array}{c}
 i\left( p+\epsilon \right)\\
0 \\
im_{0} \\
m_2 \\
\end{array}\right).
\end{aligned}
\end{equation}
Here $\epsilon = \sqrt{p^{2} + m^{2}_{\text{eff}}} $  and $m_{\text{eff}}=\sqrt{m^{2}_{0}-m^{2}_{2}}$ is the effective mass of the particle  \cite{JonesSmith2010a}. The eigenvalues corresponding to $ \widehat{v}_i$ are $-\epsilon$ and to $ \widehat{u}_i$ are $\epsilon$.
\cbl
These eigenvectors are orthogonal in our $\cPT$ inner product where  $\left< u| v \right>_\mathcal{PT} = (\mathcal{P} u)^\dagger v =  u^\dagger(-\mathbf{p}) S^\dagger v(\mathbf{p})$. 
\vskip .2cm
We  obtain  the $\cC$ operator
\be
\begin{aligned}
   \widehat{ \mathcal{C} }
    &=
    \frac{1}{\sqrt{\mathbf{p}^2+m_0^2-m_2^2}} \left(
\begin{array}{cccc}
 \mathbf{\sigma.p} & 0 & m_0 & -i m_2 \\
 0 & \mathbf{\sigma.p} & i m_2 & m_0 \\
 m_0 & i m_2 & -\mathbf{\sigma.p} & 0 \\
 -i m_2 & m_0 & 0 & -\mathbf{\sigma.p} \\
\end{array}
\right)
\end{aligned}
\ee
and the completeness relation
\begin{equation}
    |\widehat{u}_1\left))(\right( \widehat{u}_1| + |\widehat{u}_2\left))(\right( \widehat{u}_2| + |\widehat{v}_1\left))(\right( \widehat{v}_1|  + |\widehat{v}_2\left))(\right( \widehat{v}_2|={\mathbb{1}},
\end{equation}
 on using our $\mathcal{CPT}$ bra,  $ (( u| = (\mathcal{CP} u)^\dagger$.

\cm
\vskip .1cm
We demonstrate oscillation phenomena using the \cCPT~ inner product (of CS), defined in the previous section. As for the  SFDM model, we  define the flavour states as 
\begin{equation}
\label{PTosc}
    \begin{aligned}
        \left|f_{1}\right)) & = \cos{\theta}\left|  \widehat{u}_1\right)) + \sin{\theta}  \left|  \widehat{v}_1\right)), \\
                \left|f_2\right)) & =\cos{\theta}\left|  \widehat{u}_2\right)) + \sin{\theta} \left|  \widehat{v}_2\right)), \\
                \left|f_3\right)) & =-\sin{\theta} \left| \widehat{u}_1\right)) + \cos{\theta} \left| \widehat{v}_1\right)), \\
                \left|f_4\right)) & = -\sin{\theta}\left|\widehat{u}_2\right)) + \cos{\theta} \left|  \widehat{v}_2\right)).
    \end{aligned}
\end{equation}
For the (CS) \cPT~ inner product, the orthonormality and completeness relation between the $ \left|f_{i}\right))$  can be verified explicitly. This allows us to consider flavour oscillation phenomena. The time evolution of the states is  
\begin{equation}
\label{PTosc2}
    \begin{aligned}
        \left|f_{1}\right)) & = e^{-i\epsilon t}\cos \theta \left|  \widehat{u}_1\right)) + e^{i\epsilon t}\sin\theta\left|  \widehat{v}_1\right)), \\
                \left|f_2\right)) & = e^{-i\epsilon t}\cos \theta\left|  \widehat{u}_2\right)) + e^{i\epsilon t}\sin\theta\left|  \widehat{v}_2\right)), \\
                \left|f_3\right)) & =-e^{-i\epsilon t}\sin\theta\left| \widehat{u}_1\right)) + e^{i\epsilon t}\cos \theta\left| \widehat{v}_1\right)), \\
                \left|f_4\right)) & = -e^{-i\epsilon t}\sin\theta\left|\widehat{u}_2\right)) +e^{i\epsilon t} \cos \theta\left|  \widehat{v}_2\right)). \\
    \end{aligned}
\end{equation}

The transition probabilities between the flavour states are shown in table \ref{transition probability h8v}

\begin{table}[h!]
    \centering
    \begin{tabular}{|c|c|c|c|c|} \hline  
         $P$&  $\left|f'_{1}(t)\right))$&  $\left|f'_{2}(t)\right))$&  $\left|f'_{3}(t)\right))$& $\left|f'_{4}(t)\right))$\\ \hline  
         $\left|f'_{1}\right))$&  $x$&  0&  $y$& 0\\ \hline  
         $\left|f'_{2}\right))$&  0&  $x$&  0& $y$\\ \hline  
         $\left|f'_{3}\right))$&  $y$&  0&  $x$& 0\\ \hline  
         $\left|f'_{4}\right))$&  0&  $y$&  0& $x$\\ \hline 
    \end{tabular}
    \caption{transition probability of the flavour states for $H_{8v}$}
    \label{transition probability h8v}
\end{table}
 \noindent where $x = \frac{1}{4} \left(\cos (4 \theta )+2 \sin ^2(2 \theta ) \cos \left(2 \epsilon t  \right)+3\right)$ and $y = \sin ^2(2 \theta ) \sin ^2\left( \epsilon t \right)$. Since $x + y =1$, the total oscillation probability is conserved. The result is similar to the ones we obtain for SFDM. The use of the CS (\cPT~) inner product leads to the conservation of oscillation probabilities. 

\subsection{A comparison with a two flavour  Hermitian oscillation formalism}
\label{compar}

\cb From Appendix \ref{NOscill}, denoting  two flavours of neutrinos as  
$\nu_e$ and $\nu_\mu$ and the mass eigenstates as $\nu_1$ and $\nu_2$,
\be
\left( {}_{\nu_{{}_{\mu}}}^{\nu_{e}} \right) =\left( \begin{matrix}\cos \theta&\sin \theta\\ -\sin \theta&\cos \theta\end{matrix} \right)\left( {}_{\nu_{{}_{1}}}^{\nu_{2}} \right).
\ee
For a neutrino with energy $E_i$ and mass $m_i$, the time evolution of the mass eigenstate is given by
\be
\nu_{i} \left( t \right) =\exp \left( -iE_{i}t \right) \nu_{i} \left( 0 \right)
\ee
in natural units. For a relativistic neutrino, to a good approximation,
$E_{i}\approx p+\frac{m_{i}^{2}}{2p}, \; i=1,2$ where $p$ is the magnitude of the 3-momentum and $m_i$ is the mass of the neutrino. The  neutrino kets, due to mixing, satisfy
$$|\nu_{e} \left( 0 \right) \rangle =\cos \theta |\nu_{1} \left( 0 \right) \rangle +\sin \theta |\nu_{2} \left( 0 \right) \rangle $$
and 
$$|\nu_{\mu} \left( 0 \right) \rangle =-\sin \theta |\nu_{1} \left( 0 \right) \rangle +\cos \theta |\nu_{2} \left( 0 \right) \rangle .$$
At time $t$
$$|\nu_{e} \left( t \right) \rangle =\cos \theta \exp(-iE_{1}t) |\nu_{1} \left( 0 \right) \rangle +\sin \theta \exp(-iE_{2}t)|\nu_{2} \left( 0 \right) \rangle $$
and 
$$|\nu_{\mu} \left( t \right) \rangle =-\sin \theta \exp(-iE_{1}t)|\nu_{1} \left( 0 \right) \rangle +\cos \theta \exp(-iE_{2}t)|\nu_{2} \left( 0 \right) \rangle .$$
\vskip .2cm
We compare this to \eqref{PTosc} and \eqref{PTosc2}. In the $\cPT$ case we have $E_1=\epsilon$ and $E_2=-\epsilon$ and there are two copies of fermion flavours with $\epsilon$ determined by an effective mass parameter associated with the Hamiltonian. For conventional oscillations, the masses are put in by hand. For the $\cPT$ case, two dimensional parameters are allowed in the Hamiltonian in such a way that the Hamiltonian is self-adjoint with respect to an unconventional metric, which is determined by the Hamiltonian itself \footnote{\cb We stress that the models we consider are the simplest in their class and so avoid the considerable algebraic complexity of more general models, but illustrate important features of \cPT ~symmetry.\cbl}. This is a hallmark of the \cPT ~formalism. Moreover the dependence on the mixing angle of the oscillation probability  is the \emph{same} as for conventional neutrino oscillations, but the dependence on parameters with the dimension of mass is quite different. The fact that the energies of the oscillating flavours are  $\epsilon$ or $-\epsilon$ is a property of the restricted Model 8.  This observation might suggest that the \cPT ~formulation  could also describe particle-antiparticle oscillations. This is intriguing and may connect with earlier approaches to BSM physics for particle-antiparticle oscillations, which invoked Lorentz violation and decoherence (related to open systems)~\cite{Ellis:2013gca,Barenboim:2006xt}. \cPT-symmetric systems typically represent a novel class of open system with balanced loss and gain (which, in the relativistic case, can be Lorentz invariant), and so our findings may be related to these earlier BSM works (based on ideas from open systems).
\cbl
\section{Conclusions} 
We have shown explicitly that a~ $\cCPT$ inner product can be constructed for typical $\cT$-odd \emph{intrinsically} $\cPT$-symmetric Hamiltonians, including a $\cPT$-symmetric version of the Dirac equation. \cb We have Lorentz invariance in the \cPT ~Dirac equation, which however is not Dirac Hermitian.\cbl We demonstrate that such inner products open up the possibility of  unconventional flavour oscillations, which conserve probability between states.
This arises since the quantum theories  are  based on self-adjointness with respect new inner products that describe a novel form of ``physical'' theory (rather than  deformations of Hermitian ones). 
\vskip .1cm
Field theoretic versions of our quantum mechanical models would  be unlike the ones proposed to date. For $\cT$-even bosonic theories, on assuming that $\cCPT $ inner products exist for interacting \cPT~ field theories, it is known that \cite{R19,Jones:2009br,R3a}  path integrals can be defined.
It would be interesting, for investigating BSM, to extend the field theoretic treatment  to $\cT$-odd theories \cite{Jones-Smith:2009qeu}.

\section*{Acknowledgements}

We thank Nick Mavromatos and Teppei Katori for interesting discussions. L.C. is supported by the King’s-China Scholarship Council. The work of S.S. is supported  by  the EPSRC grant EP/V002821/1.

\appendix

\section{\cb Neutrino oscillations \cbl}
\label{NOscill}
\cb The study of neutrino oscillations has advanced considerably both through theoretical models \cite{RevModPhys.59.671} and through experimental observations \cite{Athar2022-ud}. Neutrinos are produced in weak interactions in one of three flavour states: electron neutrinos ($\nu_{e}$), muon neutrinos ($\mu_{e}$), or tau neutrinos ($\tau_e$). However, these flavour states are not mass eigenstates.Instead, neutrinos are superpositions of mass eigenstates $\nu_{1} ,\nu_{2} ,\nu_{3}$, which evolve over time as they propagate. This quantum mechanical phenomenon leads to neutrino oscillations, where the probability of of detecting a neutrino in one flavour state changes as a function of distance and energy.
\vskip .2cm
In Hermitian theories the probability of neutrino oscillation depends on:
\begin{itemize}
    \item The difference between the  squares of the neutrino masses $\left( \Delta m^{2} \right)$,
    \item The mixing anglesthat relate the mass and flavour eigenstates
    \item The energy of the neutrino and the distance it travels.
\end{itemize}

In the full phenomenological theory of neutrino oscillations, the three neutrino flavours are related to the three mass eigenstates by the $3 \times 3$ unitary Pontecorvo-Maki-Nakagawa-Sakata (PMNS) matrix. The parameters in the matrix are three mixing angles and one Dirac CP-violating phase. The three-flavour framework explains several key experimental results on solar, atmospheric and reactor neutrinos.
\vskip .2cm
In many experimental contexts, a simplified two-flavour model is an excellent approximation. In this framework, neutrino oscillations are described by a single mixing angle $\theta$ and a single mass squared difference $\left( \Delta m^{2} \right)$. The oscillation probability is given by:
\be
P\left( \nu_{\alpha} \rightarrow \nu_{\beta} \right) =\sin^{2} \left( 2\theta \right) \sin^{2} \left( \frac{\Delta m^{2}L}{4E} \right),
\ee
where
\begin{itemize}
    \item $L$ is the baseline distance travelled,
    \item $E$ is the neutrino energy,
    \item $\Delta m^{2}$ is the mass squared difference,
    \item $\theta$ is the mixing angle.
\end{itemize}
This approximation works well when one mass-squared difference is much larger than the others, allowing for the oscillations to be effectively two-flavoured at certain energy and distance scales.

  Current data on neutrino masses comes from a combination of cosmological observations, oscillation experiments and direct detection measurements (such as Super-Kamiokande, SNO, T2K and NOVA). These experiments do not give the absolute masses but provide precise constraints on the mass splittings between the neutrino states ( on using the theory of \emph{conventional} neutrino oscillations). Current constraints are:
  \begin{itemize}
      \item  Mass-square differences (from oscillation experiments \cite{Esteban:2020cvm,T2K:2023smv}):
       \begin{itemize}
       \item $\Delta m_{21}^{2}\approx 7.42\times 10^{-5\ }eV^{2}$,
       \item $\Delta m_{32}^{2}\approx 2.5\times 10^{-3\ }eV^{2}$.
         \end{itemize}
      \item Upper limit on individual neutrino mass (from the Karslsruhe Tritium Neutrino Experiment \cite{KATRIN:2021uub})
        \begin{itemize}
        \item $m_{\nu_{e}}<0.8eV$.
         \end{itemize}
         \item Upper limit on the sum of neutrino masses (from cosmology 
 \cite{Planck:2018vyg})
         \begin{itemize}
         \item $\sum m_{\nu}<0.12eV$.
           \end{itemize}
  \end{itemize}

  Oscillation experiments can only give information about differences of masses.

\cbl
\section{Eigenfunctions for $H_C$}
\label{HCeigen}
The solutions for the eigenvalue equation
\be
H_{C}|e_{i})=\lambda_{i} |e_{i}),\quad i=1,\ldots,4,
\ee
are \cb  $\lambda_{1}= \lambda_{2}=-1,{\rm and} \; \lambda_{3}= \lambda_{4}=1 $.\cbl  ~The corresponding eigenvectors are
\be
|e_{1})=\left( \begin{matrix}\sin \theta \  e^{-i\phi }\sin \psi \sinh (\chi /2)\\ -\frac{i}{2} \sech (\chi /2)\  \sinh\chi (\cos \psi -i\sin \psi \cos \theta )\\ 0\\ \cosh(\chi /2)\end{matrix}  \right) ,
\ee

\be
|e_{2})=\left( \begin{matrix}\frac{1}{2} \sech(\chi /2)(-i\cos \psi +\sin \psi \cos \theta \  )\sinh\chi \\ \sin \theta \  e^{i\phi }\sin \psi \  \sinh\chi /2\\ \cosh(\chi /2)\\ 0\end{matrix}  \right), 
\ee

\be
\left[ e_{3} \right) =\left( \begin{matrix}\cosh \left( \frac{\chi }{2}  \right) \sin \theta \sin \psi e^{-i\phi }\\ -\frac{1}{2} i\cosech \left( \frac{\chi }{2}  \right) \left( \cos \psi -i\cos \theta \sin \psi  \right) \sinh \chi \\ 0\\ \sinh \frac{\chi }{2} \end{matrix}  \right) ,
\ee
and
\be
\left[ e_{4} \right) =\left( \begin{matrix}\frac{1}{2} \cosech \left( \frac{\chi }{2}  \right) \left( -i\cos \psi +\cos \theta \sin \psi  \right) \sinh \chi \\ \cosh \left( \frac{\chi }{2}  \right) \sin \theta \sin \psi e^{i\phi }\\ \sinh \frac{\chi }{2} \\ 0\end{matrix}  \right) .
\ee

\section{\cb Self-adjoint condition \cbl}
\label{sec.appendix self adjointness}
\textcolor{black}{
In this appendix we examine the self-adjoint condition \eqref{PT self adjoint condition} for the two types of \cPT~  inner products (denoted by JSM and CS)  for (\cT-odd) \cPT~ symmetry; we show  both eventually lead to the same relations,
\begin{equation}
\label{condition 5}
\begin{aligned}
      -  \boldsymbol{\alpha}^\dagger  S^\dagger  &=   S^\dagger \boldsymbol{\alpha} \\
     {\rm and} \qquad \qquad \\
      \beta^\dagger S^\dagger  &=   S^\dagger \beta ,
\end{aligned}
\end{equation}
which agree with condition (v) in \cite{JonesSmith2010a}. Recall that, for an operator $H_D$ to be self-adjoint under the \cPT 
 ~inner product, it has to satisfy the condition
\begin{equation}
\label{self adjoint appendix}
    \left< \Phi \mid H_D \Psi  \right>_{\cPT} =  \left< H_D \Phi \mid  \Psi  \right>_{\cPT},
\end{equation}
where $H_D$ is the Dirac Hamiltonian given by \eqref{Dirac Hamiltonian sec 2} and $\Phi$ and $\Psi$ are some arbitrary states given by Equation \eqref{Dirac}, which depend on the momentum and position of the state; the states  are composed of two parts: the spinor part, such as $\psi(\mathbf{k})$, and the exponential plane wave part $e^{i \mathbf{k}\cdot \mathbf{r}} $. Similarly, the Dirac Hamiltonian  splits into two parts. In  the first term, $-i \boldsymbol{\alpha} \cdot \nabla$, the matrix $\boldsymbol{\alpha}$ acts on the spinor parts, while the differential operator $\nabla$ acts on the exponential part. Hence, 
\begin{equation}
    \begin{aligned}
    -i \boldsymbol{\alpha} \cdot \nabla  \Psi(\mathbf{k},\mathbf{r}) =& \boldsymbol{\alpha}  \cdot \mathbf{k} \Psi(\mathbf{k},\mathbf{r}) .
    \end{aligned}
\end{equation}
Therefore the Hamiltonian becomes an operator that depends on the momentum of the state on which it acts.
}

\subsection{\cb JSM inner product \cbl}
\cb
We provide a more detailed derivation of \eqref{condition 5} for the original JSM inner product, given in \cite{JonesSmith2010a}; this  facilitates an appreciation of  the difference between the two inner products.
From the definition of the JSM inner product
\begin{equation}
    \left< \phi \mid \psi  \right>_{\cPT}  =\  (\mathcal{PT}\phi )^{T} Z  \psi,
\end{equation}
on substituting the explicit form of $\Phi$ and $\Psi$, the left hand side of Eq. \eqref{self adjoint appendix} becomes
\begin{equation}
    \begin{aligned}
         \left< \Phi \mid H_D \Psi  \right>_{\cPT} =& (\mathcal{PT}\Phi (\mathbf{k} ,\mathbf{r}) )^{T} Z H_D (\mathbf{p}) \Psi(\mathbf{p} ,\mathbf{r})  \\
         =& \int_\mathbf{r} (\cPT u(\mathbf{k}) e^{i \mathbf{k} \cdot \mathbf{r}})^\top Z H_D(\mathbf{p}) \, v(\mathbf{p}) e^{i \mathbf{p} \cdot \mathbf{r}}).
    \end{aligned}
\end{equation}
Since the states involve two parts, the inner product can  be viewed as the composition of the contributions from the spinor part and the plane wave part. The spinor part of the inner product follows from the matrix representation of the operators \cP ~and \cT~ on the states.  The plane wave part involves the inner product in position space, given by an integration over  $x$ from $-\infty$ to $\infty$. Apart from a matrix representation, the parity operator $\cP$ also reverses the sign of momentum $\mathbf{k}$ and position $\mathbf{r}$, and the time-reversal operator \cT  ~reverses the sign of momentum $\mathbf{k}$ and $i$ due to the canonical commutation relation\footnote{
Time-reversal flips the sign of the commutator $[x,p]=i\hbar$, and so the $i$ on right hand side has to change to $-i$.}. Hence the combined action of \cP~  and \cT  ~changes the spinor part from $u(\mathbf{k})$ to $u^*(-\mathbf{k})$, while leaving the plane wave part unchanged. So we find

\begin{equation}
    \begin{aligned}
         \left< \Phi \mid H_D \Psi  \right>_{\cPT}
         =& \int_\mathbf{r} (S Z u^*(\mathbf{-k}) e^{i \mathbf{k} \cdot \mathbf{r}})^\top Z H_D(\mathbf{p}) \, v(\mathbf{p}) e^{i \mathbf{p} \cdot \mathbf{r}}\\
           =& \int_\mathbf{r} (Z S^*u^*(\mathbf{-k}) e^{i \mathbf{k} \cdot \mathbf{r}})^\top Z H_D(\mathbf{p}) \, v(\mathbf{p}) e^{i \mathbf{p} \cdot \mathbf{r}}\\
            =& \int_\mathbf{r} u^\dagger(\mathbf{-k}) e^{i \mathbf{k} \cdot \mathbf{r}} S^\dagger Z^\top Z H_D(\mathbf{p}) \, v(\mathbf{p}) e^{i \mathbf{p} \cdot \mathbf{r}}\\
              =& \int_\mathbf{r} e^{i (\mathbf{k}+\mathbf{p}) \cdot \mathbf{r}} u^\dagger(\mathbf{-k})  S^\dagger H_D(\mathbf{p}) \, v(\mathbf{p}) \\
                =&  (2\pi)^3 \delta(\mathbf{k}+\mathbf{p})  u^\dagger(\mathbf{-k})  S^\dagger H_D(\mathbf{p}) \, v(\mathbf{p}). 
    \end{aligned}
\end{equation}
We have used that \cP  ~and \cT  
  ~commute 
 
  and $Z Z^\top = 1$ for $\cT^2 =-1$ in the basis of \eqref{canonical z}. Similarly, the right hand side of equation \eqref{self adjoint appendix} becomes 
\begin{equation}
    \begin{aligned}
        \left< H_D \Phi \mid  \Psi  \right>_{\cPT} =& (\mathcal{PT} H_D (\mathbf{k}) \Phi (\mathbf{k} ,\mathbf{r}) )^{T} Z  \Psi(\mathbf{p} ,\mathbf{r})  \\
         =& \int_\mathbf{r} (\cPT H_D(\mathbf{p}) u(\mathbf{k}) e^{i \mathbf{k} \cdot \mathbf{r}})^\top Z  \, v(\mathbf{p}) e^{i \mathbf{p} \cdot \mathbf{r}})\\
          =& \int_\mathbf{r} (S Z H^*_D(\mathbf{k}) u^*(\mathbf{-k}) e^{i \mathbf{k} \cdot \mathbf{r}})^\top Z  \, v(\mathbf{p}) e^{i \mathbf{p} \cdot \mathbf{r}}\\
             =& \int_\mathbf{r} u^\dagger(\mathbf{-k}) e^{i \mathbf{k} \cdot \mathbf{r}}H^\dagger_D(\mathbf{k}) S^\dagger Z^\top Z  \, v(\mathbf{p}) e^{i \mathbf{p} \cdot \mathbf{r}}\\
           =&  (2\pi)^3 \delta(\mathbf{k}+\mathbf{p})  u^\dagger(\mathbf{-k}) H^\dagger_D(\mathbf{k})  S^\dagger  \, v(\mathbf{p}) .
    \end{aligned}
\end{equation}
Since the delta function $\delta(\mathbf{k}+\mathbf{p})$ is none zero only when $\mathbf{k}=-\mathbf{p}$, combining the LHS and the RHS gives
\begin{equation}
\label{JSM final derivation}
    \begin{aligned}
         u^\dagger(\mathbf{p})  S^\dagger H_D(\mathbf{p}) \, v(\mathbf{p}) =&    u^\dagger(\mathbf{p}) H^\dagger_D(\mathbf{-p})  S^\dagger  \, v(\mathbf{p}) ,\\
        S^\dagger H_D(\mathbf{p}) =& H^\dagger_D(\mathbf{-p})  S^\dagger.
    \end{aligned}
\end{equation}
On using the explicit expression for the Dirac Hamiltonian \eqref{Dirac Hamiltonian sec 2}, we obtain \eqref{condition 5}.

\subsection{CS inner product}

We examine \eqref{self adjoint appendix} using the CS inner product: $\left( \phi \mid \psi  \right)_{\cPT}  =\  (\mathcal{P} \phi )^{\dagger}  \psi. $ 
The parity operator \cP  ~just acts on the ket state $\phi$,  
since it leaves the exponential part of the state unchanged. On substituting explicit forms of $\Psi$ and $\Phi$, the left hand side of Eq. \eqref{self adjoint appendix} becomes
\begin{equation}
    \begin{aligned}
           \left( \Phi \mid H_D \Psi  \right)_{\cPT} =& 
           (\mathcal{P}\Phi (\mathbf{k} ,\mathbf{r}) )^{\dagger} H_D (\mathbf{p}) \Psi(\mathbf{p} ,\mathbf{r})  \\
              =& \int_\mathbf{r} (\cP u(\mathbf{k}) e^{i \mathbf{k} \cdot \mathbf{r}})^\dagger H_D(\mathbf{p}) \, v(\mathbf{p}) e^{i \mathbf{p} \cdot \mathbf{r}})\\
              =& \int_\mathbf{r} (S u(\mathbf{-k}) e^{i \mathbf{k} \cdot \mathbf{r}})^\dagger H_D(\mathbf{p}) \, v(\mathbf{p}) e^{i \mathbf{p} \cdot \mathbf{r}}\\
               =& \int_\mathbf{r} u^\dagger(\mathbf{-k}) e^{-i \mathbf{k} \cdot \mathbf{r}} S^\dagger H_D(\mathbf{p}) \, v(\mathbf{p}) e^{i \mathbf{p} \cdot \mathbf{r}}\\
               =&  (2\pi)^3 \delta(\mathbf{k}-\mathbf{p})  u^\dagger(\mathbf{-k})  S^\dagger H_D(\mathbf{p}) \, v(\mathbf{p}) .
    \end{aligned}
\end{equation}
Similarly, the right hand side of Eq.\eqref{self adjoint appendix}  becomes
\begin{equation}
    \begin{aligned}
           \left( \Phi H_D \mid \Psi  \right)_{\cPT} =& 
           (\mathcal{P} H_D(\mathbf{k})\Phi (\mathbf{k} ,\mathbf{r}) )^{\dagger}  \Psi(\mathbf{p} ,\mathbf{r})  \\
           =& \int_\mathbf{r} (\cP  H_D(\mathbf{k}) u(\mathbf{k}) e^{i \mathbf{k} \cdot \mathbf{r}})^\dagger  \, v(\mathbf{p}) e^{i \mathbf{p} \cdot \mathbf{r}})\\
            =& \int_\mathbf{r} (S H_D(-\mathbf{k}) u(\mathbf{-k}) e^{i \mathbf{k} \cdot \mathbf{r}})^\dagger  \, v(\mathbf{p}) e^{i \mathbf{p} \cdot \mathbf{r}}\\
             =& \int_\mathbf{r} u^\dagger(\mathbf{-k}) e^{-i \mathbf{k} \cdot \mathbf{r}}  H^\dagger_D(-\mathbf{k})  S^\dagger\, v(\mathbf{p}) e^{i \mathbf{p} \cdot \mathbf{r}}\\
             =&  (2\pi)^3 \delta(\mathbf{k}-\mathbf{p})  u^\dagger(\mathbf{-k}) H^\dagger_D(-\mathbf{k})  S^\dagger \, v(\mathbf{p}) .
    \end{aligned}
\end{equation}
Hence \; $u^\dagger(\mathbf{-p})  S^\dagger H_D(\mathbf{p}) \, v(\mathbf{p}) =    u^\dagger(\mathbf{-p}) H^\dagger_D(-\mathbf{p})  S^\dagger \, v(\mathbf{p})$

and so $ S^\dagger H_D(\mathbf{p}) = H^\dagger_D(-\mathbf{p})  S^\dagger$.

\vskip .2cm
\noindent Finally, we deduce  Eq. \eqref{JSM final derivation}, which shows that Eq. \eqref{condition 5} follows from Eq. \eqref{self adjoint appendix}  (using the CS inner product).

\cbl

\bibliography{bibliPTDirac.bib}

\end{document}